# Electron spectroscopy of single quantum objects to directly correlate the local structure to their electronic transport and optical properties

AUTHOR NAMES

Ryosuke Senga,[1] Thomas Pichler[2] and Kazu Suenaga[1]*

AUTHER ADDRESS

[1]*Nano-Materials Research Institute, National Institute of Advanced Industrial Science and Technology (AIST), Tsukuba 305-8565, Japan.*

[2]*Faculty of Physics, University of Vienna, Strudlhofgasse 4, A-1090 Vienna, Austria*

*Correspondence to: suenaga-kazu@aist.go.jp

ABSTRACT

   Physical property of a single quantum object is governed by its precise atomic arrangement. The direct correlation of localized physical properties with the atomic structures has been therefore strongly desired but still limited in the theoretical studies. Here, we have successfully examined the localized electronic properties of individual carbon nanotubes by means of high-resolution electron energy-loss spectroscopy combined with high-resolution transmission electron microscopy. Well-separated sharp peaks at the carbon $K$(1s) absorption edge and in the valence-loss spectra are obtained from a single freestanding carbon nanotube with the local chiral index and unambiguously identified as the transitions between the van Hove singularities. The spectra features clearly vary upon the different areas even in the individual carbon nanotube. Variations in interband transitions, plasmonic behaviors and unoccupied electronic structures are clearly attributed to the local irregular atomic arrangement such as topological defect and/or elastic bond stretching.

**Main text**

The diversity in the electronic structure of single quantum object depending on its atomic arrangement is the most fascinating aspect of material physics. The carbon nanotube[1], though made up of a single element, has been theoretically predicted to exhibit a wide range of electronic properties from metallic to semiconducting based on its chirality[2,3]. However, it is not feasible to date to directly correlate the distinct electronic structure involving a van Hove singularity (vHs) with the exact chiral index (n, m) of an individual single-walled carbon nanotube (SWNT) especially with irregular atomic arrangements. For instance, direct measurements of the occupied and unoccupied density of states (DOS) by photoemission and X-ray absorption (XAS)[4–9] do not have the spatial resolution to identify the local changes of the band structure on a single SWNT. Single SWNT experiments using scanning tunneling spectroscopy (STS) allow a direct comparison of the local DOS but are complicated by modifications due to substrate screening[10]. On the other hand, for the excited state properties, as measured in optical experiments, excitonic effects are crucial in comparing to the joint DOS[10].

Electron energy-loss spectroscopy (EELS) combined with a transmission electron microscope (TEM) can investigate the DOS of an isolated single quantum object with an atomic resolution in both core-loss and valence-loss spectroscopy. EELS was actually used to distinguish individual fullerene molecules as different carbon allotropes[11]. TEM provides a full picture of the carbon network arrangement[12,13]. However, the energy resolution of TEM-based EELS was not sufficient to identify the vHs in SWNTs because the energy spread of a cold field-emission gun widely used as an electron source is approximately 0.3-0.5 eV. We therefore employed a more sophisticated electron source with a monochromator consisting of double Wien-filter and improved the instability to achieve the energy resolution to 24 meV in the zero-loss peak at the full width at half maximum (FWHM)[14]. To visualize the detailed atomic structure (involving any defect or chirality multiplicity) of a targeted SWNT, two delta-type high-order aberration correctors[15] were mounted on the monochromated TEM.

Figure 1 shows a TEM image and the low energy-loss (low-loss) and core-level energy-loss (core-loss) spectra from an isolated SWNT obtained by our monochromated TEM. By inserting energy selecting slits in the dispersion plane of the monochromator, the energy resolution can be practically adjusted stepwise from 30 to 200 meV (Fig. 1b). In the core-loss spectrum (C1s) obtained with the energy slit (4-µm) inserted (solid line in Fig. 1c), multiple peaks in the π* response are clearly resolved, which cannot be observed in the spectrum taken at non-monochromated condition (without any energy slit shown as broken line in Fig. 1c). Low-loss spectra were also obtained for the same SWNT at a higher energy resolution (30–100 meV with 0.25–1.3-µm slits). A typical low-loss spectrum has several peaks (Fig. 1d). These peaks indicate the interband transitions between the vH singularities in the valence and conduction bands. Note that core-loss and low-loss spectra were calibrated by unsaturated zero-loss peaks simultaneously recorded by the dual EELS mode. We also assign the atomic structure of the SWNT, such as the chirality or possible defects, from the simultaneously recorded high-resolution scanning TEM (STEM)/TEM images (Fig. 1a). The chiral index of the examined SWNT can be easily determined by a fast Fourier transform (FFT) pattern (inset) following Zhu *et al.* [13], in which the effect of the tube tilt with respect to the projection plane can be minimized.

In this manner, we have analyzed several tens of SWNTs with different chiralities and categorized the spectra as metallic SWNTs ($2n + m = 3k$) or semiconducting SWNTs ($2n + m \neq 3k$). Note that defective SWNTs were excluded from the series, and only nondefective, freestanding, and clean (contamination free) SWNTs that are isolated from other particles are shown. The diameters of these SWNTs widely range from 0.75 to 1.22 nm.

For the core-loss spectra as shown in Fig. 2a and 2b, every SWNT has multiple peaks at 285 eV (solid red lines) in the π* response (1s → π*), which cannot be observed in the graphene π* response (broken black line). The π* fine structures were previously demonstrated by XAS using bundled SWNTs[6,7,9] and assigned to localized vHs in the unoccupied DOS by comparing to diameter cumulative tight-binding (TB) calculations[6,7]. The delocalized graphitic states were assigned to the overall broad π* resonance which is downshifted by about 2eV due to core-hole effects. More recently, XAS on bulk samples of chirality mixture of (6, 5) and (6, 4) tubes confirmed this assignment by *ab-initio* calculations[8]. However the π* fine structural analysis for individual SWNTs has never been performed, except for calculations[16].

In order to corroborate the fine structure on an individual SWNT basis, we employed a similar line shape analysis following the reports[6–8]. The line shape analysis reveals four peaks related to the vHs ( $E_1^*$ to $E_4^*$) in the (6, 5) DOS and a broad π* resonance. Comparing to the shifted *ab-initio* DOS[8] (inset in Fig. 2c), one finds a good matching with the simulated spectrum regarding the vHs position. Disagreement in the relative intensities can be due to the core-hole effects, which are not included in the model. These core-hole effects also lead to additional lifetime broadening of about 0.1 eV in the FWHM. We compare our (6, 5) spectrum with the XAS data in Ref. 8 obtained from (6, 5) and (6, 4) mixed SWNT bundles (Fig. S1). They are very similar but slightly different because of the additional fine structure related to (6, 4) tubes present in the sample. This implicates the importance of single SWNT experiments to unambiguously assign the unoccupied DOS from core level EELS. Unfortunately no *ab-initio* calculations are available except for (6, 5) SWNTs, therefore a first and less accurate comparison with TB calculations was performed here (Fig. S1).

As expected the σ* peak exhibits less variety depending on the chirality. As the diameter of the SWNT increases, the peaks increase sharply, which is attributed to the curvature effect of SWNTs on the σ* features[17].

Turning to an analysis of the valence band excitations. The EELS below 25 eV are covered by inter and intra-band excitations governed by the joint DOS and strongly excitonic. The interband excitations are at zero momentum transfer directly related to optical interband transitions via a Kramers-Kronig transformation of the loss function. For the low-loss spectra, several peaks appear around 1–5 eV which have been assigned to localized non-dispersive interband transitions between the vHs[18]. In contrast intra-band plasmon excitations of free charge carriers and the overall π and π+σ plasmon are strongly dispersive with increasing momentum transfer[18].

The spatially resolved valence band EELS spectra for the individual metallic and semiconducting SWNTs are depicted in Figs 3a and 3b, respectively. The peaks can be consequently assigned either to the interband transitions between the vHs in the valence and conduction bands for each SWNTs or to a free charge carrier plasmon in the metallic tubes. Though our measurements are integrating over the SWNT Brillouine zone, the localized nature of these transitions allows a direct comparison to optical experiments taking into account that the peak plasmon positions are slightly above the absorption maxima. The corresponding values of $E_{ii}$ (i = 1, 2, …) for semiconducting SWNTs are in good agreement with empirical Kataura plot based on the optical measurements (Fig. 3c)[19], except for SWNTs with low chiral angles (Fig. S2). The values of $M_{ii}$ (i = 1, 2, …) for metallic SWNTs are also in good accordance with the optical measurements in literature. Splitting of the $M_{11}$ peaks ($M_{11}^+$ and $M_{11}^-$)—probably caused by the trigonal wrapping effect[20,21]—is observed. Further studies will allow us go beyond to determine and normalize the loss function to the sum rule extracting the full two particle excitation spectra including correlation effects.

In addition a rather large peak around 1 eV (marked by asterisk*) is observed in the loss function for metallic tubes. These peaks never appeared in semiconducting SWNTs and contributions from TEM grid or substrate are excluded as shown in Fig. S3. They can be therefore assigned to the free charge carriers metallic SWNTs. The peak energy is higher than one expected from the simple Drude model at zero momentum transfer. In addition, the literatures report a wide range of plasmon energies from a few tens of meV up to 1-2 eV of SWNTs for undoped[22,23] and doped bundles[24–26] by optical and EELS measurements. Hence, the effect that we measure at an average momentum transfer and/or unintended doping from the TEM supporting grid might explain this high plasmon energy. This explanation is further supported by the fact that the plasmon peak located in the near-infrared range is only observed in isolated metallic SWNTs and apparently diminishes in bundled SWNTs (Fig. S3) and is reduced at defect sites (Fig. 5). However this, the EELS is a direct analogy to the optical measurements in such "surface objects", because the response function has basically to be described as a generalized polarizability for localized excitations[27,28]. In this model, any charge carrier plasmon should be silent. Therefore, in order to describe the 1 eV peak in the metallic nanotubes, additional depolarization effects have to be taken into account. Indeed, momentum dependent studies on aligned samples of individual SWNTs revealed the importance of local field corrections for delocalized states polarized along the tube axis[29]. For our momentum integrated studies, such local field corrections could yield to depolarization effects due to a mixing of higher momentum states into the generalized polarizability and additional peaks which cannot directly be compared to optics. This especially accounts for a high energy shoulder in the π plasmon and for the 1D charge carrier plasmon. Thus the dissipative low energy contribution in metallic SWNTs can be tentatively assigned to a charge carrier plasmon polarized along the tube axis. A further study with the various doping level will be required to fully unveil its origin and extract the optical conductivity on an absolute scale as its position is not directly related to the electron density due to the local field corrections.

A great benefit of using the C $K$-edge as a fingerprint of the local electronic properties of SWNTs is that the information obtained from the core-loss spectra is quite localized so that even the two closely packed SWNTs in Fig. 4a can be distinguished. The 1s to π* energy-loss near-edge structure (ELNES) with multiple peaks can abruptly change from tube to tube (Fig. 4b), whereas the valence-loss spectra involving the vHs indeed show considerable mixing (Fig. 4c). Therefore, tube-by-tube identification is possible by core-loss but not by valence-loss. When two SWNTs maintain a certain distance from each other (5–6 nm apart), as shown in Fig. 4d, the identification of both SWNTs is also possible from the valence-loss spectra (Fig. 4e). However, even in this case, the intensities of $E_{ii}$ or $M_{ii}$ are widely spread up to a few nanometers, even the outside the SWNTs (Fig. 4f). The different degree of delocalization of the low- and core-loss signals shown here is related to the spatial extension of the evanescent coulombic field associated with the induced excitations[30–33]

Despite the delocalization of the low-loss signals, no striking coupling mode was detected in the optical range (Fig. 4c). On the other hand, valence region measurements obtained by bulk EELS or local DOS obtained by STS should be more carefully interpreted because the spectrum of an SWNT can be largely affected by neighboring tubes, impurity particles, or substrates if any. A completely isolated SWNT (5–6 nm apart), which is freestanding, will be most suitable to be analyzed as a single quantum object.

A major advantage of EELS performed in TEM compared to other methods is the capability of local spectroscopy of non-periodic structures. In particular, the interrupted periodicities of SWNTs or the perturbations induced by an external field can largely alter their electronic properties, and the spatial distribution of defect states will be of primary importance in future device applications. The commercially available SWNT specimens used in this experiment often contain many defects due to both plastic and elastic deformation (Fig. 5 and Figs. S6, S7) or serial junctions with multiple chiralities (Fig. 6).

We have found that both plastic and elastic deformation of a SWNT can induce local electronic structure changes. Figure 5 shows plastically deformed SWNTs (with topological defects). The SWNT has the same chirality (12, 3) above and below the defective point. Indeed, the C K-edge fine structure taken from both sides of the defective point exhibits completely identical spectra. However, at the defective point, the C K-edge exhibits a tremendous change in the 1s → π* intensity, and both π* and σ* become broad (purple line in Fig. 5c). On the other hand, the low-loss spectra recorded from the same position do not show any obvious change in the $M_{ii}$ peaks among spectra ii and iii (Fig. 5d). In other words, the plasmonic interband transitions are hardly affected at the defective area. This is in complete agreement with the degree of delocalization of these excitonic plasmonic electron-hole interband transitions introduced above.

In contrast, the peak around 0.9 eV which is assigned to a strongly dispersive delocalized plasmon of the free charge carriers is considerably shifted at the defective area to about 0.6 eV (Fig. 5d). The plasmon energy position at defect sites can be directly related to the square root of the relative density of the free electron like charge carriers. Therefore, the charge carrier density is only about 40 % at the defective area pointing towards a significant degree of localization of the electrons. This can be an origin of the inferior conductivity of defective nanotubes. Such a local measurement of the free electron like plasmonic behavior highlights the future possibility to quantify the modification in the conductivity of an individual nanotube in the presence of identified defects and directly correlate it to their electronic transport.

On the other hand, the elastically bent SWNT (Fig. S7) does not exhibit such obvious changes in the $\pi^*$ or $\sigma^*$ peak shapes, but the peak positions are systematically shifted when the spectra obtained at the inside and outside of the bend of a SWNT are compared. Note that a simple mechanical consideration suggests a compressive stress at the inside of the bent SWNT while there is a tensile stress for the outside. The peaks in the $\pi^*$ feature for the compressed region are located at an energy that is approximately 100 meV lower than those for the tensile region, whereas the $\sigma^*$ exhibits a non-negligible upshift on the compressed side. These shifts presumably reflect the quite small increase in the bond length (less than a few percent, as estimated from the bending angle in Fig. S7). It is noteworthy that such unique changes induced by the bond length stretching are only observed in the core-loss spectra and not in the low-loss spectra.

A hybrid SWNT serial junction consisting of SWNTs with more than two different chiralities can be regarded as a simple nanodevice such as a diode, but its distinct local electronic structure has not been thoroughly investigated. A TEM image of typical hybrid SWNT is shown in Fig. 6a. This hybrid SWNT involves a serial junction between the thinner (11, 1) semiconducting part and the thicker (10, 10) metallic part, as shown in the FFT patterns. The low-loss spectra of each SWNT taken at a position ~5 nm away from the junction indicate $E_{ii}$ (i = 1–3) and $M_{ii}$ (i = 1, 2) to confirm the semiconducting and metallic nature of both SWNTs, respectively (Fig. 6b). It is noteworthy that the plasmon peak discussed above can only be observed for the metallic SWNT. A series of core-loss spectra around the junction point (i to x) is presented in Figs. 6d and 6e, which correspond to the dotted lines (i) to (x) across the SWNT in Fig. 6c. Interestingly, the π* peaks in the junction, shown as the green lines in Fig. 6e, are different form the spectra for either (10, 10) and (11, 1). In this case, a broader peak is not observed in both the π* and σ* features, unlike the simple defects shown in Fig. 5. There are new peaks in the 1s → π* response in spectra iv to vii indicated by the black arrows. This implies that the junction has a distinct electronic structure due to the two connected SWNTs.

The present results involve future perspective such as direct correlation of the two particle excitation spectra determined in valence band excitations and the matrix element weighted unoccupied DOS measured by core level EELS with the local atomic structure. Further EELS measurements with momentum resolution could even verify the dispersive relationship of 1D quantum object, which is, however, not in the scope of this letter. Nevertheless, this method will have unprecedented implication for the exact determination of the influence of the environment (e.g. neighboring tubes or substrates), and local structures including non-periodic structures (e.g. topological defects or serial junctions) on the electronic and optical properties. This will be a crucial step in the basic description of the correlated electronic structure in these nanomaterials as it will allow to directly extract for instance spatially resolved exciton binding energies, the size extension of electron-hole and electron-electron correlations and so on. Finally, such localized electronic properties of atomically defined SWNTs would accelerate the realization of single-molecule devices such as biosensors or field-emission transistors fabricated with well-defined SWNT segments.

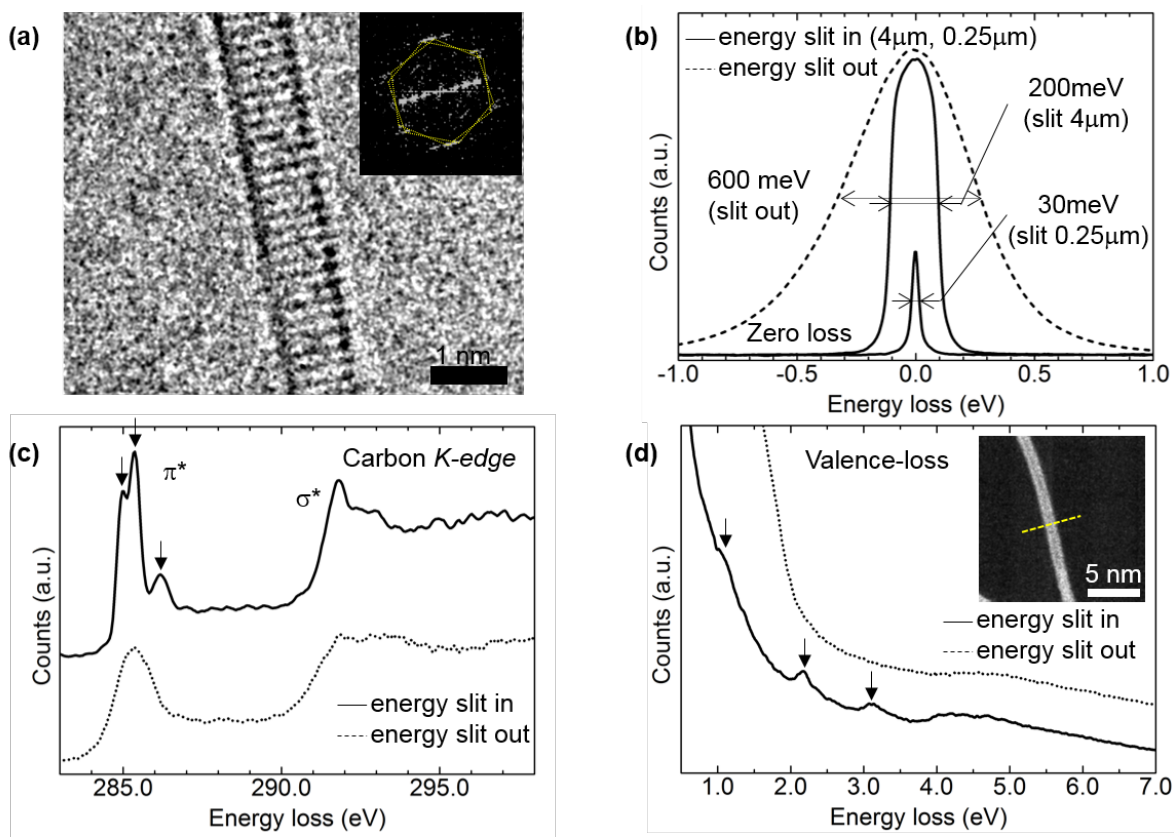

**Figure 1.** High-resolution imaging and high-resolution spectroscopy of an individual SWNT. (a) A TEM image of the examined SWNT and its FFT pattern (inset) showing its chiral index as (10, 1). (b) Energy resolutions of 600 meV (without the slit), 200 meV (4-µm slit inserted), and 30 meV (0.25-µm slit inserted), measured by zero-loss peak at FHWM. (c) The C *K*-edge (C1s) spectrum recorded when the electron probe scanned across the yellow line (inset of (d)), showing the well-separated peaks in π* region with the energy slit inserted (solid line, 200-meV resolution). (d) Low-loss spectrum recorded with the smaller energy slit (solid line, 30-meV resolution). The SWNT is freestanding and well isolated from other tubes or impurity particles to avoid their contributions in the low-loss energy region.

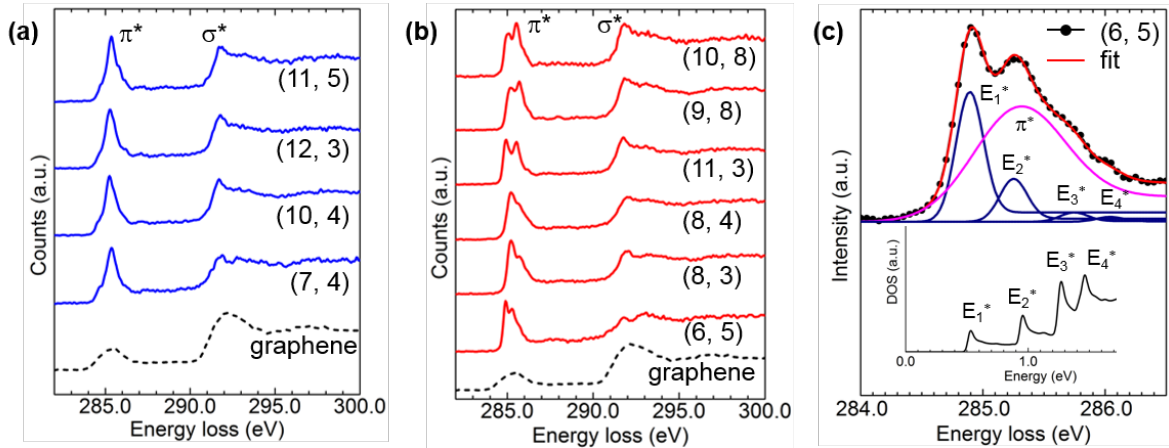

**Figure 2.** The chirality-dependent fine structure in the unoccupied DOS. (a),(b) The core-loss spectra for metallic and semiconducting SWNTs, respectively. A spectrum for graphene is also shown as a reference (black dotted lines at the bottom). (c) The fine structures of the π* regions for (6, 5) with the results of a line shape analysis and peak assignment to the overall π* resonance (pink solid line) and the vHs (purple lines). The lower inset presents the *ab initio* calculated DOS from Ref. 8. Core-hole effects must be considered for more detailed analysis so that the C1s response allows a direct comparison to the unoccupied DOS.

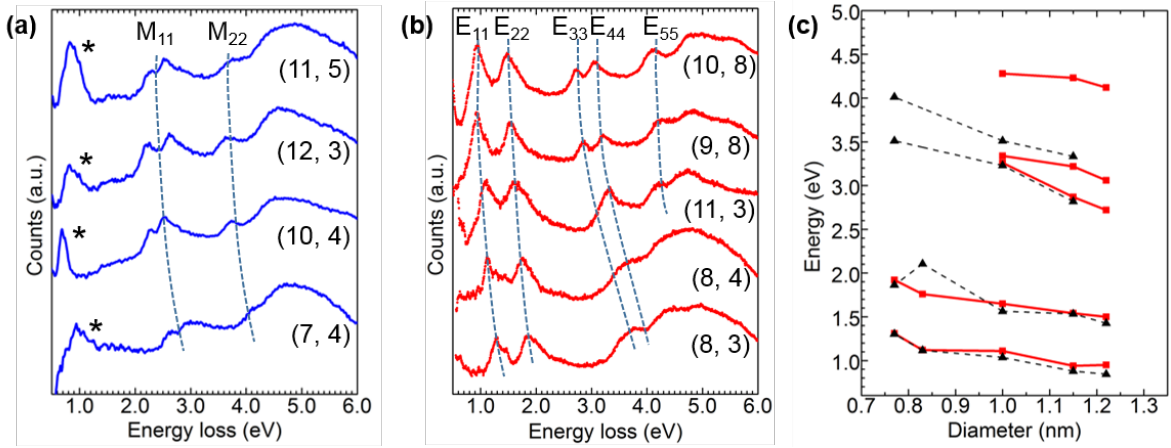

**Figure 3.** Chirality dependence of the valence-loss spectra. (a),(b) The low-loss spectra for metallic and semiconducting SWNTs, respectively. The peaks in the low-loss region are assigned as the interband transitions ($M_{ii}$ and $E_{ii}$, dotted lines) and exhibit monotonic downshifts as the diameter increases. (c) Diameter dependence of the interband transitions for semiconducting SWNTs measured from our experiment (red solid lines) showing a good agreement with existing optical measurements[19] (black broken lines). The possible origin of the peaks around 1 eV in (a) denoted by * is the plasmonic feature of metallic SWNTs. All data for the SWNTs in the graph are listed in Table S1.

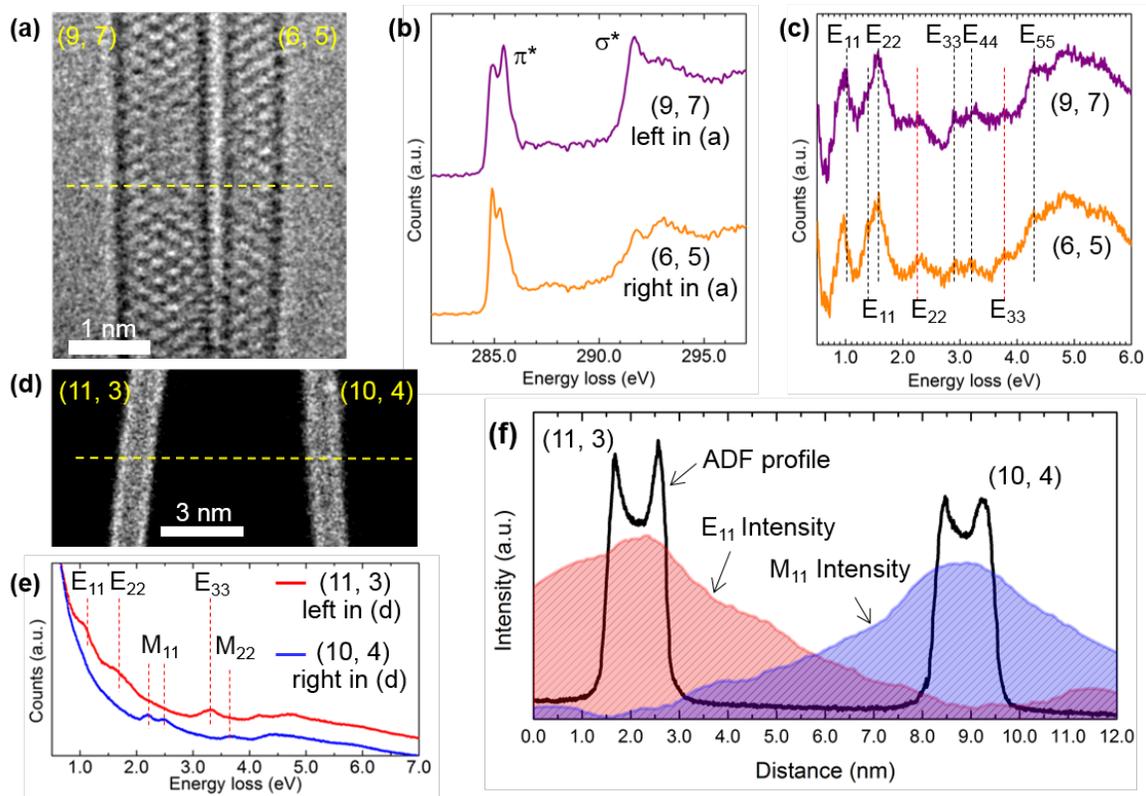

**Figure 4.** Delocalization of low-loss and core-loss spectra between two adjacent SWNTs. (a) STEM image of two closely aligned SWNTs identified as (9, 7) for the left and (6, 5) for the right. Although the core-loss spectra in (b) clearly show the different features between the two SWNTs, the low-loss spectra in (c) show almost the same features, in which the vH singularity peaks for both SWNTs are mixed and unable to be spatially discriminated. (d) STEM image of two SWNTs separated by more than 5 nm. The left and right SWNTs are assigned as (11, 3) and (10, 4) by other TEM images (not shown), respectively. (e) Low-loss spectra for the left (red line) and right (blue line) SWNTs picked up from a spectrum line recorded when the electron probe scanned along the yellow line in (d). (f) The distribution of the $E_{11}$ peak intensity of the left SWNT (red line) and the $M_{11}$ peak intensity of the right SWNT (blue line) with the simultaneously recorded ADF profile (black line), showing a very large delocalization (FWHM of 5–6 nm). More detailed information is described in Figs. S4 and S5.

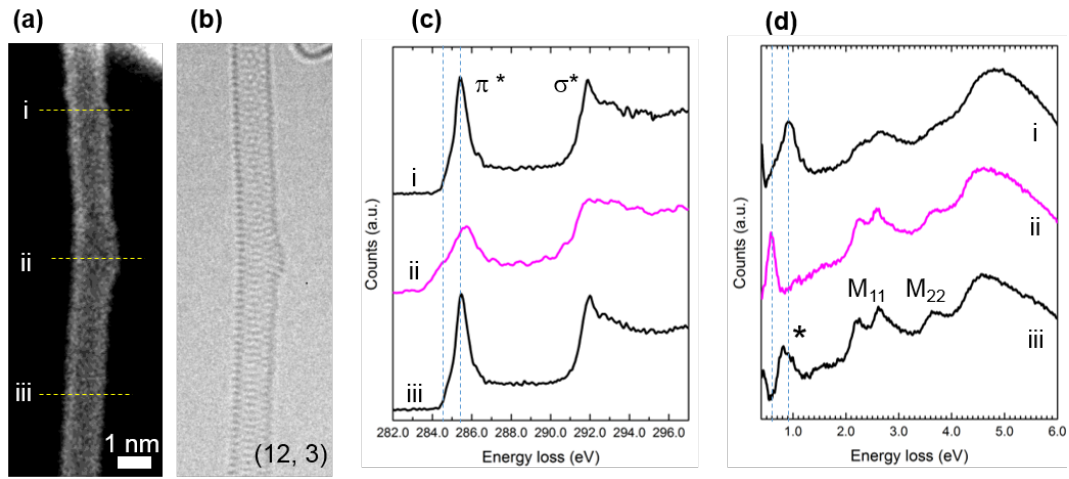

**Figure 5.** EELS spectra for an SWNT with topological defects. (a),(b) STEM and TEM images, respectively, of an SWNT that has a bump consisting of topological defects on the wall. (c),(d) The C *K*-edge and low-loss spectra recorded when the electron probe scanned across the yellow lines denoted by i, ii, and iii in (a). The broad peaks of $M_{11}$ and $M_{22}$ in spectra i is caused by the existence of a large substance observed at the right upper corner of (a) and (b).

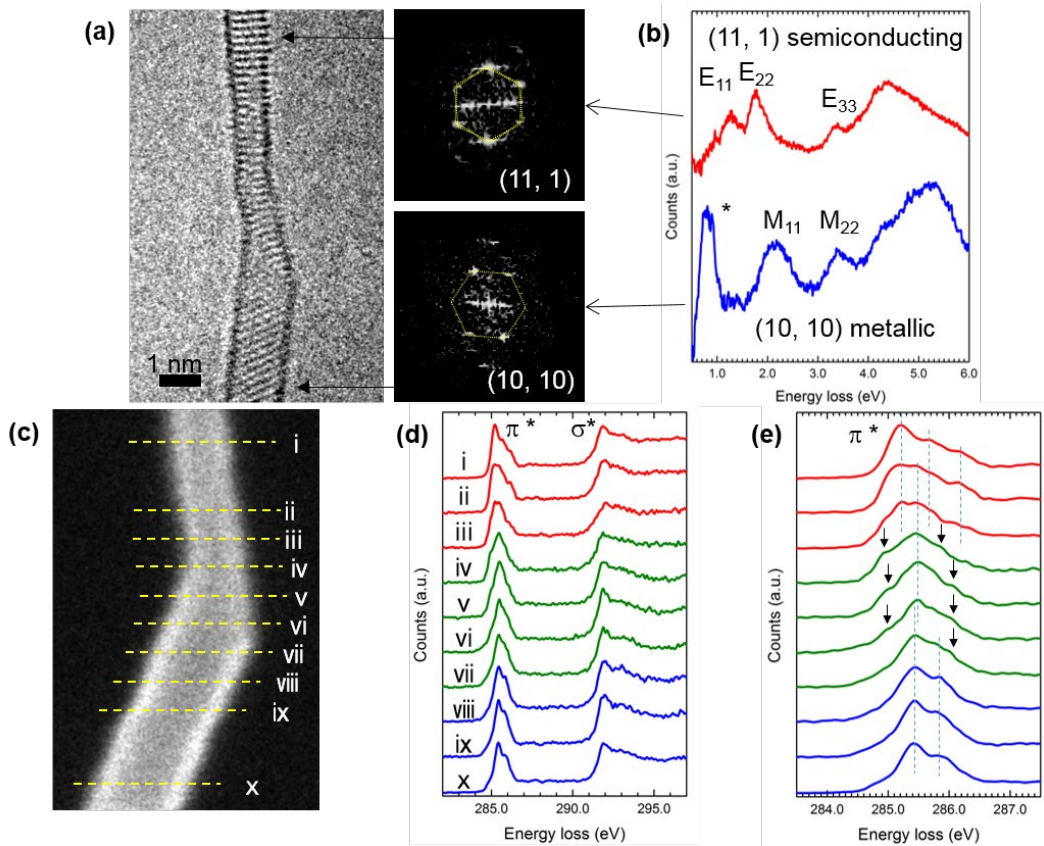

**Figure 6.** EELS spectra for a hybrid SWNT serial junction consisting of metallic and semiconducting SWNTs. (a) TEM image of the junction between (11, 1) and (10, 10) nanotubes. The chiralities are identified from the FFT patterns (inset). (b) The low-loss spectra of both SWNTs recorded far enough (more than 5 nm away) from the junction. (c) STEM image of the same SWNT junction. The C $K$-edge shown in (d) and (e) is recorded when the electron probe is scanned across the yellow lines denoted by i to x in (c).

## ASSOCIATED CONTENT

**Supporting Information**

The Supporting Information is available free of charge on the ACS Publications website at DOI:

Materials and Methods. Comparison with XAS data and tight binding (TB) calculations (Figure S1). Low-loss and core-loss spectra for low-chiral-angle CNTs (Figure S2). Valence-loss spectra for an isolated SWNT and bundled SWNTs (Figure S3). High-resolution spectroscopy of two closely packed SWNTs (Figure S4). Delocalization of vH singularity peaks in EELS (Figure S5). EELS spectra for an SWNT with a kink (Figure S6). EELS spectra for an elastically deformed SWNT (Figure S7). Atomic structure and measured $E_{ii}$ values for the SWNTs mentioned in the main text (Table S1).

## AUTHOR INFORMATION

**Corresponding Author**

E-mail: suenaga-kazu@aist.go.jp

**Author contributions**

RS and KS designed experiments. RS prepared materials, performed microscopy. RS and TP analyzed data. TP performed theoretical calculations. RS, TP and KS co-wrote the paper.

**Notes**

The authors declare no competing financial interests.

## ACKNOWLEDGEMENTS

We thank Dr. Sawada, Dr. Sasaki, Dr. Mukai and Dr. Morishita with JEOL Co., Ltd. for the development of our new monochromated TEM (TripleC#2). We also acknowledge Dr. Maruyama, Dr. Kataura, Dr. Okazaki, Dr. Nakanishi, and Dr. Yanagi for fruitful discussions. This work was supported by the Japan Society for the Promotion of Science (JSPS) and the JST Research Acceleration Program. T.P. thanks the FWF P27769-N20 for funding.

Insert Table of Contents Graphic and Synopsis Here

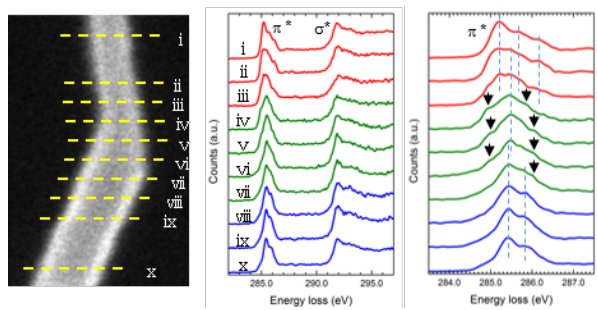

# Supplementary information for

# Electron spectroscopy of single quantum objects to directly correlate the local structure to their electronic transport and optical properties

**Authors:** Ryosuke Senga,[1] Thomas Pichler[2] and Kazu Suenaga[1*]

**Affiliations:**

[1] *Nano-Materials Research Institute, National Institute of Advanced Industrial Science and Technology (AIST), AIST Central 5, Tsukuba 305-8565, Japan.*

[2]*Faculty of Physics, University of Vienna, Strudlhofgasse 4, A-1090 Vienna, Austria*

*Correspondence to: suenaga-kazu@aist.go.jp

**This PDF file includes:**

Materials and Methods

Figures S1 to S7

Table S1

**Materials and Methods:**

We have used a JEOL TEM (3C2) equipped with a Schottky field emission gun, a double Wien filter monochromator (Ref. [14] in the main text) and delta correctors (Ref. [15] in the main text). In the monochromator, the energy resolution can be practically adjusted stepwise from 30 to 200 meV by inserting energy selecting slits in the dispersion plane of the monochromator. Note that an energy resolution of approximately 150 meV is sufficiently trustworthy to identify the vH singularity at the C *K*(1s) edge for any SWNT, whereas a higher energy resolution of approximately 25–30 meV is required for the vH singularity in the low-loss region. EELS data have been collected with an improved Gatan Quantum spectrometer designed for low-voltage operation with a higher stability. This unique apparatus enables us to obtain high-energy-resolution and high-resolution imaging simultaneously at low-voltage operation (15–60 kV). The STEM imaging and EELS experiments presented here were performed at 60 keV. The convergence semiangle and the EELS collection semiangle were 42.5 mrad and 33 mrad, respectively. The probe current was 33 pA for core-loss spectroscopy. The energy dispersion of the spectrometer was set to 5 and 50 meV/channel for low-loss and core-loss spectroscopy, respectively. We used commercially available HiPCO SWNTs that were dispersed onto grids and heated in vacuum at 200 °C for 12 h to prevent contamination.

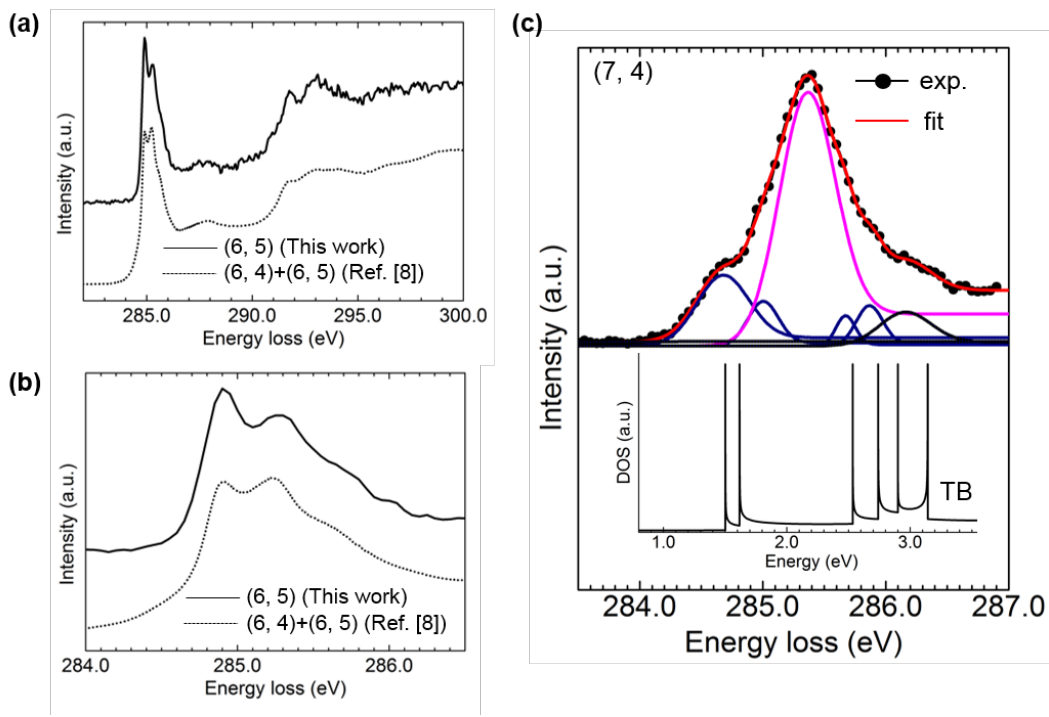

**Figure S1| Comparison with XAS data and tight binding (TB) calculations.** (a),(b) C1s fine structures for (6, 5) (solid line) and (6, 4)/(6, 5) mixture obtained by XAS from Ref. [8] in the main text (broken line). Both spectra are in a good overall agreement. (c) The fine structures of the $\pi^*$ regions for (7, 3) with the results of a line shape analysis. The experimentally obtained spectra are indicated by the black lines and the solid circles. Each fitting line (red) includes the Gaussian peaks corresponding to the vHs (purple) and a broad Voigtian peak corresponding to the $\pi^*$ resonance (pink). DOS for each SWNT are calculated by tight binding (TB from Ref. [7] in the main text) approximation as shown in the inset. The overall agreement is fairly good although some deviations regarding the positions and number of peaks point suggest the need of more accurate ab-initio calculations including core-hole effects.

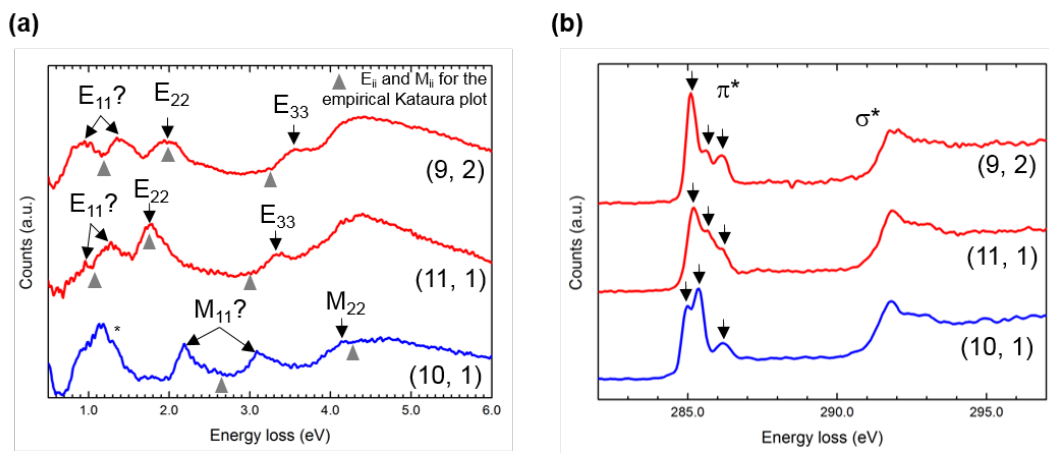

**Figure S2| Low-loss and core-loss spectra for low-chiral-angle CNTs.** (a) Low-loss spectra and (b) core-loss spectra for (9, 2), (11,1), and (10, 1) SWNTs; the chiral angles are 9.82°, 4.31°, and 4.72°, respectively. The near-zigzag SWNTs exhibit considerable deviations in the peak positions of $E_{ii}$ and $M_{ii}$ expected from the empirical Kataura plot (Ref. [19] in the main text), as indicated by the gray triangles in (a). In the semiconducting SWNTs, unidentified peaks appear around $E_{11}$, which appear to be split from the $E_{11}$ peaks, as is the case for the $M_{11}$ peaks of the metallic SWNTs. The second and third peaks for the (10, 1) SWNT are also not fully consistent with the theory and difficult to assign in the scheme of the simple zone-folding picture. In order to fully explain these unidentifiable peaks, further theoretical studies are necessary.

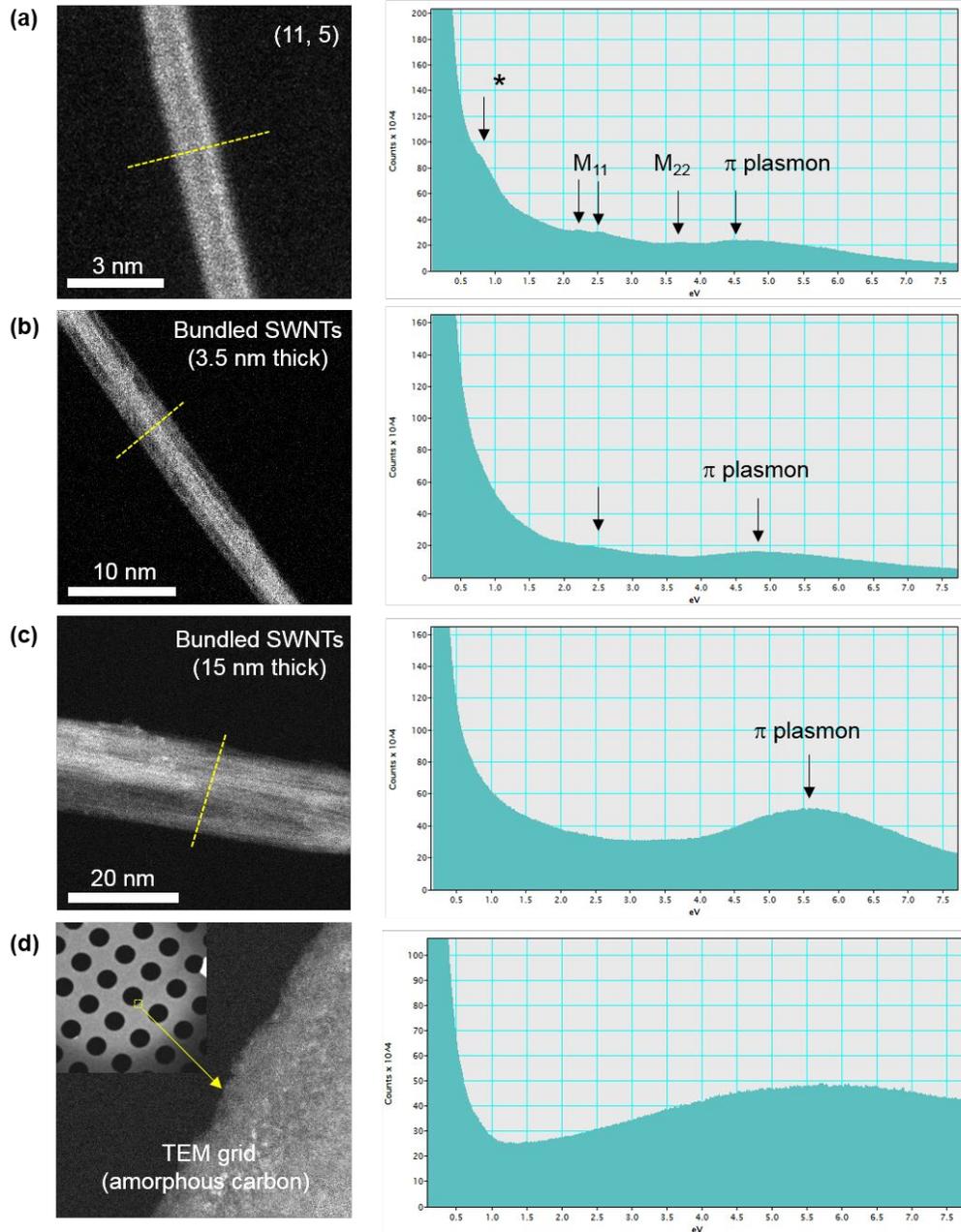

**Figure S3| Valence-loss spectra for an isolated SWNT and bundled SWNTs.** (a) STEM image (left panel) and low-loss spectrum (right panel) for an isolated (11, 5) metallic nanotube. The plasmonic feature is clearly observed around 1 eV (indicated by *) along with vH singularities around 2–4 eV. (b),(c) STEM images (left panel) and low-loss spectra (right panel) for the bundled SWNTs. The thicknesses of the bundles are 3.5 nm for (b) and 15 nm (c). In (b), the plasmonic feature cannot be observed, whereas the peaks correlated with the vH singularities are slightly observed. In the much thicker bundle shown in (c), the peaks related to the plasmon and vH singularities cannot be observed, probably owing to a large $\pi \rightarrow \pi^*$ plasmon that appeared in the range of 4–6 eV. (d) The energy loss from the TEM grid (amorphous carbon) showing no sharp peak around 1 eV.

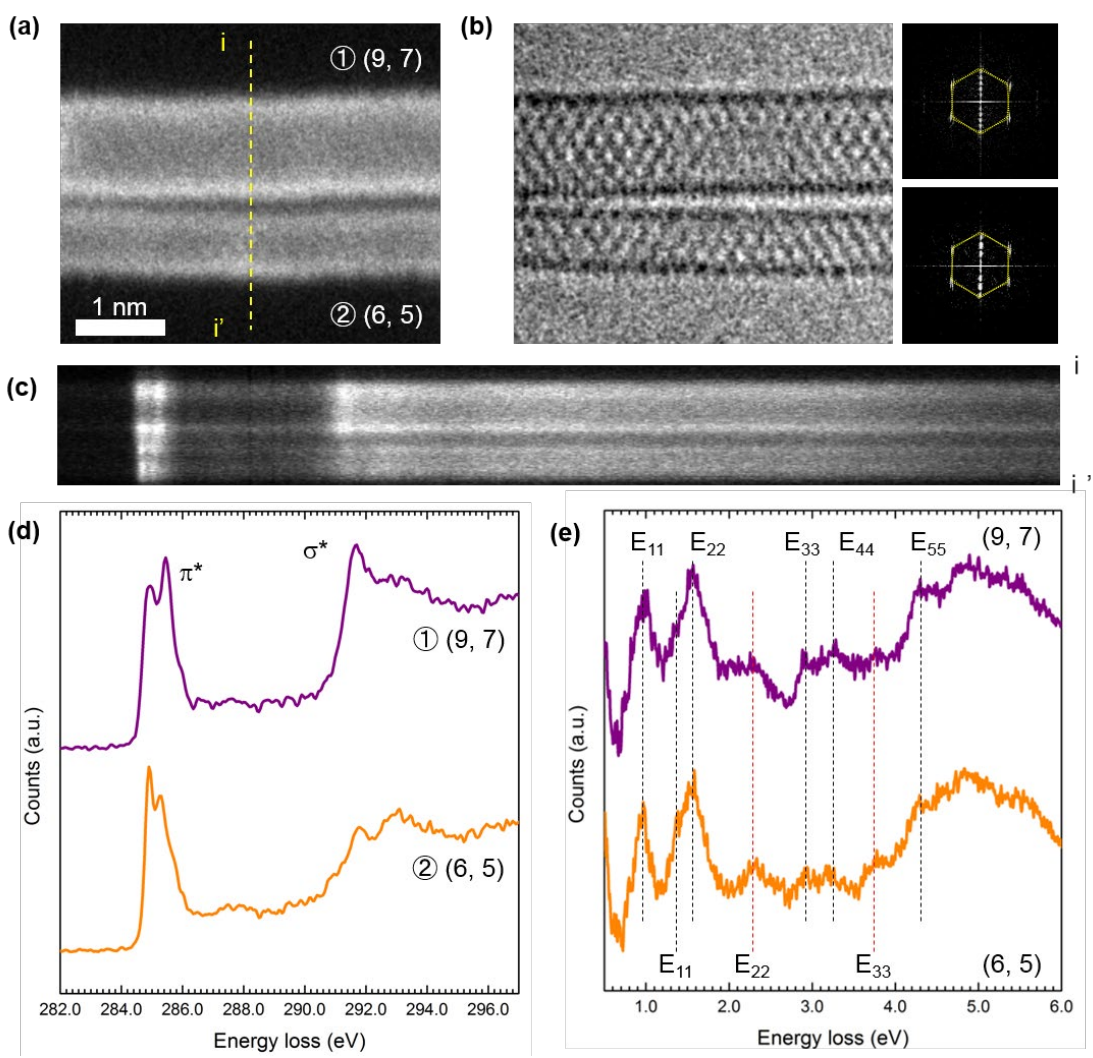

**Figure S4| High-resolution spectroscopy of two closely packed SWNTs.** (a),(b) STEM and TEM images of two aligned SWNTs. The thicker SWNT (top) and the thinner one (bottom) are assigned as (9, 7) and (6, 5), respectively, from the FFT patterns for each (right panels in (b)). (c) The spectrum image near the C *K*-edge recorded when the electron probe scanned across the yellow line (i–i') in (a). Each spectrum in (d) shows completely different features between the two examined SWNTs. Although the π* response has sharp double peaks in both cases, the peak positions and intensities are different. In addition, the σ* peak for the (6, 5) SWNT is much broader than the (9, 7) SWNTs. On the other hand, the low-loss spectra taken from same two SWNTs do not show any differences between them. The spectra contain $E_{ii}$ for both SWNTs and cannot be distinguished.

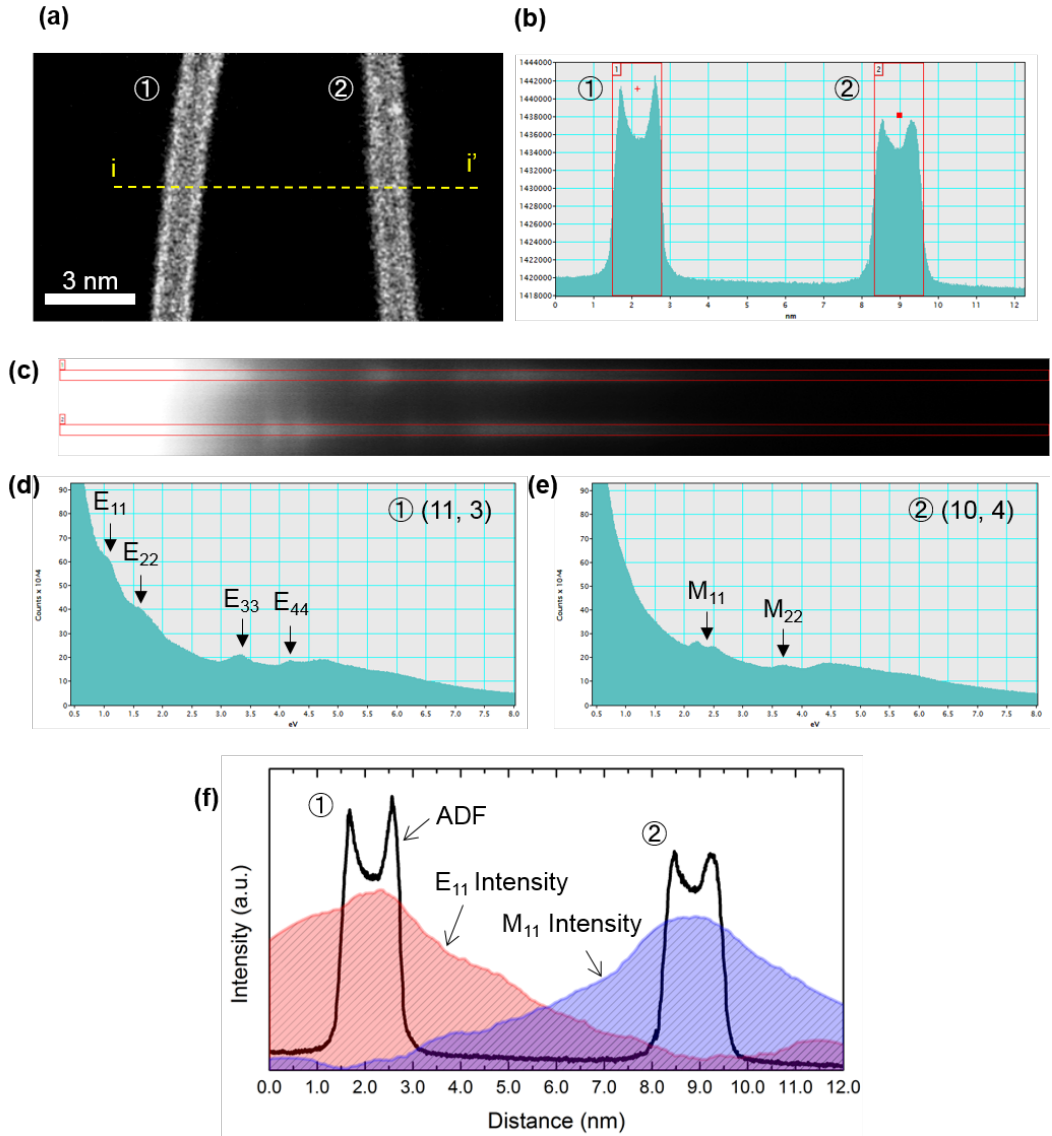

**Figure S5| Delocalization of vH singularity peaks in EELS.** (a) An STEM-ADF image of two SWNTs. The SWNTs are well separated at a distance of ~7 nm from each other. **c**, The spectrum image of the valence region recorded when the electron probe scanned across the yellow line (i–i') in (a). The corresponding ADF profile is shown in (b). The left and right SWNTs are assigned as (11, 3) and (10, 4), respectively, from their FFT patterns (not shown). Note that we obtained TEM and STEM images to assign these SWNTs because the well separated and freestanding SWNTs often exhibit self-vibration, and it is difficult to image the atomic configurations in STEM mode. (d),(e) The low-loss spectra for each SWNT that sum the spectra in the red squares in (c). The left SWNT (11, 3) and the right SWNT (10, 3) exhibit $E_{ii}$ (i = 1–4) and $M_{ii}$ (i = 1–2) peaks, respectively. (f) The distribution of the $E_{11}$ peak intensity of the left SWNT (red line) and the $M_{11}$ peak intensity of the right SWNT (blue line) are integrated in an energy window of 0.5 eV after subtraction of the power-law background and then presented with the simultaneously recorded ADF profile (black line), showing a very large delocalization (FWHM of 5–6 nm). Note that the relatively lower intensities for $E_{11}$ and $M_{11}$ for the other nanotubes are probably caused by dielectric screening from the counterpart nanotubes.

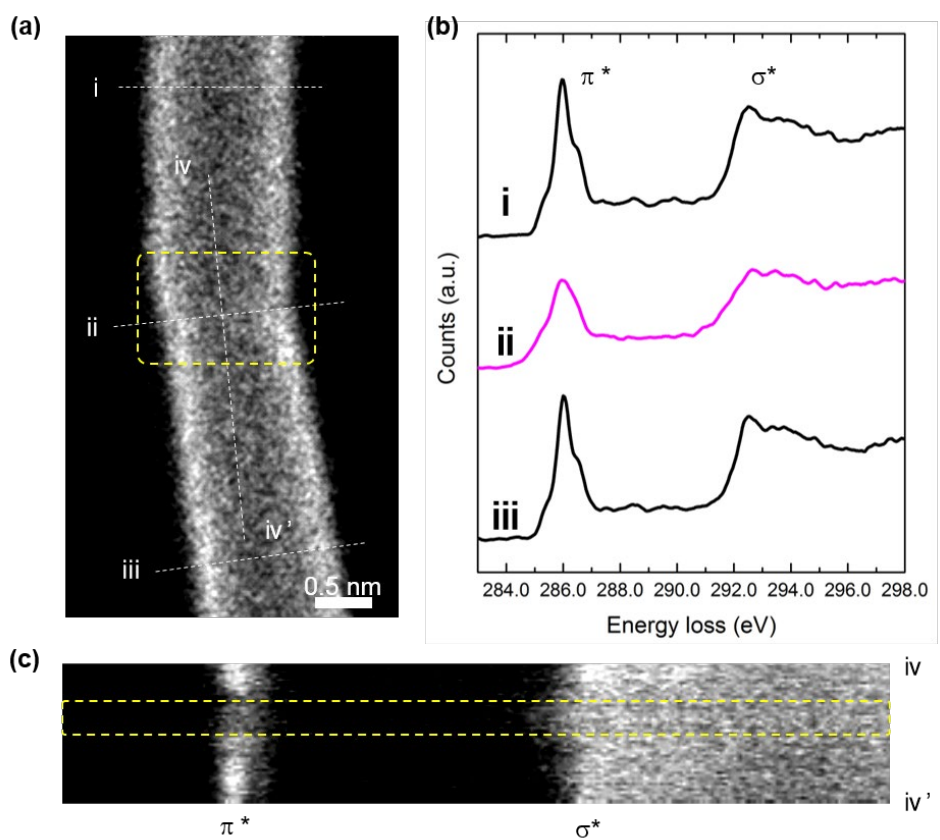

**Figure S6| EELS spectra for an SWNT with a kink.** (a) An STEM image of a mechanically bent SWNT. The SWNT has a single chirality throughout its axis but is bent at one point, probably owing to an external force induced between two fixed points on the specimen grid. (b) The C $K$-edge fine structure taken from both sides of the bent region, showing completely identical subpeaks (marked by the arrows), which demonstrates the same chirality for both sides. Interestingly, the carbon $K$-edge at the kink position exhibits a drastic change in the $\pi^*$ intensity, and both $\pi^*$ and $\sigma^*$ become broad, as is the case for the SWNT with topological defects in Fig. S5. (c) The line spectrum during a line scan along the tube axis (iv to iv') showing the localized states in the bending area within 1.2 nm.

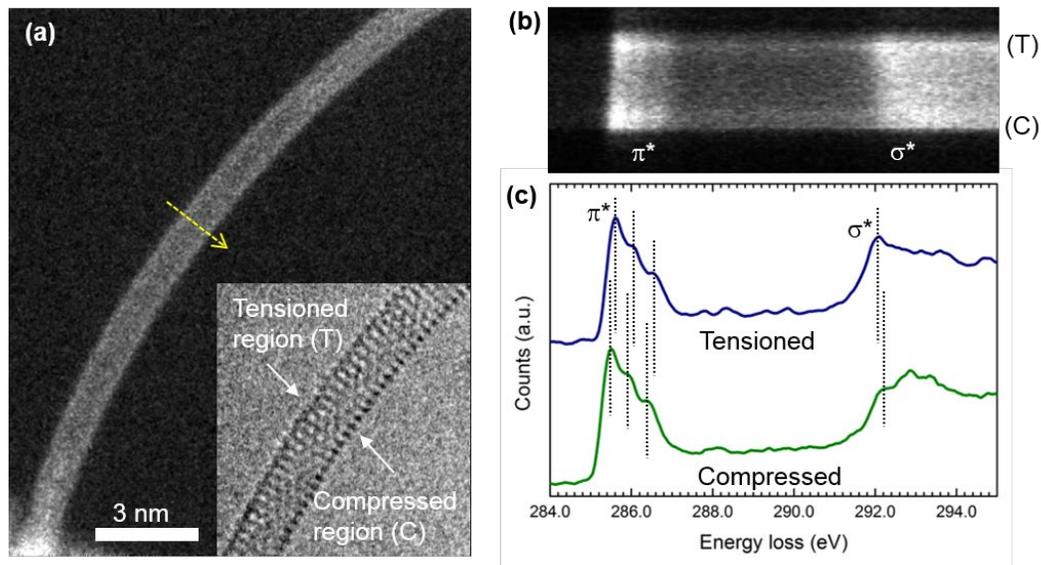

**Figure S7| EELS spectra for an elastically deformed SWNT.** (a) STEM and TEM (inset) images of an elastically bent SWNT. The bending curvature is roughly estimated as 0.05 (1/nm) from a low-magnification image. (b) The core-loss spectra images recorded when the electron probe is scanned along the yellow arrows in (a). When the scan starts on the compressed side (inside of the curve) following arrow in (a), the π* peak downshifts as the electron probe moves to the tensioned side (outside of the curve), whereas the σ* peak slightly upshifts, as shown in (b). Note that the spectral image is calibrated by zero-loss peak positions which are simultaneously recorded. Therefore the peak shifts are not resulted by scanning of the beam. (d) The spectra obtained from the tensioned side (blue line) and the compressed side (green) show a systematic energy shift of ~100 meV. In addition to the peaks shifts, the σ* peak exhibits a considerably broader structure on the compressed side. Such changes are probably induced by a small increase in the bond length due to the elastic deformation. For the SWNT in (a), the bond stretching can be estimated as a few percent on the tensioned and compressed sides from the bending curvature.

|  | Chirality | | Diameter (nm) | Chiral angle (°) | van Hove singularity (eV) | | | | | | |
|---|---|---|---|---|---|---|---|---|---|---|---|
|  | n | m |  |  | $E_{11}$ | $E_{22}$ | $M_{11}^{-}$ | $M_{11}^{+}$ | $E_{33}$ | $E_{44}$ | $M_{22}$ |
| Semiconducting SWNTs | 8 | 3 | 0.77 | 15.30 | 1.31 (1.30) | 1.92 (1.86) |  |  | - | - |  |
|  | 9 | 2 | 0.79 | 9.83 | 0.95? (1.09) | 1.93? (2.25) |  |  | - | - |  |
|  | 8 | 4 | 0.83 | 19.11 | 1.12 (1.11) | 1.76 (2.10) |  |  | - | - |  |
|  | 11 | 1 | 0.90 | 4.31 | 1.40? (0.98) | 1.76 (2.03) |  |  | 3.47 (3.09) | - |  |
|  | 11 | 3 | 1.00 | 11.74 | 1.11 (1.04) | 1.65 (1.56) |  |  | 3.26 (3.23) | 3.34 (3.51) |  |
|  | 9 | 8 | 1.15 | 28.05 | 0.94 (0.88) | 1.54 (1.53) |  |  | 2.87 (2.81) | 3.22 (3.33) |  |
|  | 10 | 8 | 1.22 | 26.33 | 0.95 (0.84) | 1.50 (1.43) |  |  | 2.72 (2.59) | 3.06 (3.04) |  |
| Metallic SWNTs | 7 | 4 | 0.76 | 21.05 |  |  | 2.60 | 2.89 (2.86) |  |  | 4.10 |
|  | 10 | 1 | 0.82 | 4.72 |  |  | 2.18 | 3.13 (2.68) |  |  | - |
|  | 10 | 4 | 0.98 | 16.10 |  |  | 2.30 | 2.52 (2.37) |  |  | 3.76 |
|  | 12 | 3 | 1.08 | 10.89 |  |  | 2.22 | 2.62 (2.21) |  |  | 3.63 |
|  | 11 | 5 | 1.11 | 17.78 |  |  | 2.24 | 2.50 (2.16) |  |  | 3.72 |
|  | 10 | 10 | 1.36 | 30.00 |  |  | 2.14 | 2.14 (1.86) |  |  | 3.40 |

**Table S1 Atomic structure and measured $E_{ii}$ values for the SWNTs mentioned in the main text.** The values in parentheses for van Hove singularity are taken from ref. [19] in the main text.